\documentclass[9pt,twocolumn,twoside]{pnas-new}
\templatetype{pnasresearcharticle} 
\newcommand{\earth}{\oplus}

\newcommand{\beginSI}{%
        \setcounter{table}{0}
        \renewcommand{\thetable}{S\arabic{table}}%
        \setcounter{equation}{0}
        \renewcommand{\theequation}{S\arabic{equation}}%
        \setcounter{figure}{0}
        \renewcommand{\thefigure}{S\arabic{figure}}%
         }

\title{LAMOST telescope reveals that Neptunian cousins of hot Jupiters are mostly single offspring of stars that are rich in heavy elements}

\author[a,1, 2]{Subo Dong}
\author[b,1, 2]{Ji-Wei Xie} 
\author[b]{Ji-Lin Zhou}
\author[c]{Zheng Zheng}
\author[d]{Ali Luo}

\affil[a]{Kavli Institute for Astronomy and Astrophysics, Peking University, Yi He Yuan Road 5, Hai Dian District, Beijing 100871, China}
\affil[b]{School of Astronomy and Space Science \& Key Laboratory of Modern Astronomy and Astrophysics in Ministry of Education, Nanjing University, 210093, China}
\affil[c]{Department of Physics and Astronomy, University of Utah, 115 South 1400 East, Salt Lake City, UT 84112, USA}
\affil[d]{National Astronomical Observatories, Chinese Academy of Sciences, Beijing 100012, China}

\leadauthor{Dong} 

\significancestatement{Hot Jupiters are Jupiter-size planets at $\lesssim1/10$ of the Sun-Earth distance, and even though they were the first exoplanet population discovered around sun-like stars, their origins still remain elusive.
Using data from NASA's {\it Kepler} satellite and China's Large Sky Area Multi-Object Fiber Spectroscopic Telescope (LAMOST) telescope, we discover a population of close-in Neptune-size planets (called ``Hoptunes'') that share key similarities with hot Jupiters. Like hot Jupiters, Hoptunes prefer to reside around stars with higher metal abundance than the Sun. Nearly half of the  {\it Kepler} planets are discovered in systems with multiple transiting planets, but both hot Jupiters and Hoptunes are preferentially found in single transiting-planet systems. The ``kinship'' between hot Jupiters and Hoptunes suggest likely common origins and offer fresh clues into the formation of these exotic close-in planets.}

\authorcontributions{S.D. and J.-W.X. designed and performed the research; S.D., J.-W.X. and A.L. analyzed the data; S.D., J.-W.X., J.-L. Zhou and Z. Zheng wrote the paper.}
\authordeclaration{The authors declare no conflict of interest.}
\equalauthors{\textsuperscript{1}S.D. and J.-W.X. contributed equally to this work.}
\correspondingauthor{\textsuperscript{2}To whom correspondence should be addressed. E-mail: dongsubo@pku.edu.cn or jwxie@nju.edu.cn }


\begin{abstract}
We discover a population of short-period, Neptune-size planets sharing key similarities with hot Jupiters: both populations are preferentially hosted by metal-rich stars, and both are preferentially found in {\it Kepler}  systems with single transiting planets.
We use accurate LAMOST DR4 stellar parameters for main-sequence stars to study the distributions of short-period $(1\,{\rm d}<P<10\,{\rm d})$ {\it Kepler} planets as a function of host star metallicity. The radius distribution of planets around metal-rich stars is more ``puffed up'' as compared to that around metal-poor hosts. In two period-radius regimes, planets preferentially reside around metal-rich stars, while there are hardly any planets around metal-poor stars. One is the well-known hot Jupiters, and the other is a population of Neptune-size planets ($2 R_\oplus \lesssim R_p \lesssim 6 R_\oplus$), dubbed as ``Hoptunes''. Also like hot Jupiters, Hoptunes occur more frequently in systems with single transiting planets though the fraction of Hoptunes occurring in multiples is larger than that of hot Jupiters. About $1\%$ of solar-type stars host ``Hoptunes'', and the frequencies of Hoptunes and hot Jupiters increase with consistent trends as a function of [Fe/H]. In the planet radius distribution, hot Jupiters and Hoptunes are separated by a ``valley'' at approximately Saturn size (in the range of $6 R_\oplus \lesssim R_p \lesssim 10 R_\oplus$), and this ``hot-Saturn valley'' represents approximately an order-of-magnitude decrease in planet frequency compared to hot Jupiters and Hoptunes. The empirical ``kinship'' between Hoptunes and hot Jupiters suggests  likely common processes (migration and/or formation) responsible for their existence.
\end{abstract}

\dates{This manuscript was compiled on \today}
\doi{\url{www.pnas.org/cgi/doi/10.1073/pnas.XXXXXXXXXX}}

\begin{document}

\verticaladjustment{-2pt}

\maketitle
\thispagestyle{firststyle}
\ifthenelse{\boolean{shortarticle}}{\ifthenelse{\boolean{singlecolumn}}{\abscontentformatted}{\abscontent}}{}

\dropcap{M}ore than two decades after the first surprising discovery \cite{mayor95}, hot Jupiters still remain to be one of the most hotly studied exoplanet populations. 
Observationally, they never seem to fail to yield new surprises. It was realized early on that their host stars were predominantly more metal rich than the sun \cite{hotjupitermet}, and their frequency was later found to strongly correlate with host [Fe/H] \cite{santos04, fischer05}. 
Lately, {\it Kepler} data show that they stand out as a distinctly ``lonely'' population for the dearth of other planets on nearby orbits in their systems \cite{lathamhotj,lonely,warmhot}. Despite the plethora of observational findings, we still do not know their origins with certainty -- we do not know how they migrate to their present close-in orbits $(P\lesssim 10\,{\rm d})$  \cite[see review by][]{review} or if they have migrated at all \cite{insitu, insitu3}.

More clues about their origins may come from examining them in the context of planet distributions and the dependence of such distributions on host environment: How unique are the conditions forming hot Jupiters? And do hot Jupiters have ``relatives'', which share similarities in their planetary and host-star properties? 

With thousands of planets discovered from monitoring $\sim 200,000$ target stars, the {\it Kepler} mission has unprecedented potential to study planet distributions over a wide range of parameter space and their possible links to stellar properties \cite[see e.g.,][]{howard12, dongzhu, fressin, pnas2013, Bur15, wangfischer, zhu16}. However, making any reliable statistical inference with a large {\it Kepler} sample is seriously limited by the lack of accurate stellar parameters for the majority of the targets. For most of the {\it Kepler} targets, stellar parameters are only available via the Kepler Input Catalog (KIC; \cite{KIC}), whose [Fe/H] and $\log g$ measurements are known to have large uncertainties and serious systematic errors \cite[see, e.g.][]{dong14, peter, xie16, california}. Significant efforts have been put into characterizing planet hosts by taking high-resolution spectroscopy \cite[e.g.,][]{buc, buc2, california} or extracting asteroseismic parameters \cite[e.g.,][]{huber}. But the accurate parameters of the underlying parent sample (with and without detected planets) are poorly known, which presents a major uncertainty in {\it Kepler} planet statistics \cite[see relevant discussions in][]{dongzhu, Bur15} and also makes it difficult to reliably derive planet distributions as a function of stellar parameters.

With 4000 fibers and $5^\circ$ diameter field of view, the 4m Large Sky Area Multi-Object Fiber Spectroscopic Telescope (LAMOST, a.k.a. Goushoujing Telescope) \cite{wang96, sucui04, cui, zhao} is uniquely positioned to perform a systematic spectroscopic survey of {\it Kepler} target stars. The ``LAMOST-{\it Kepler} project'' \cite{decat, ren} attempts to observe all {\it Kepler} target stars with no preference for known planet hosts. It forms a large and unbiased sample to perform statistical inference on planet distributions and correlations with host properties \cite{dong14}.  For main-sequence stars, the stellar parameters inferred from the official LASP (LAMOST Stellar Parameter Pipeline) pipeline \cite{luo} are demonstrated to be accurate (typical uncertainties: $\sigma{T_{\rm eff}} = 100 K$, $\sigma{\log g} = 0.1-0.15$\,dex, $\sigma{\rm [Fe/H]} = 0.1$\,dex) from comparisons with high-resolution spectroscopic and asteroseismic parameters (see Section 1 of Supporting Information of \cite{xie16} and also \cite{dong14, wang16}.) The accurate LASP stellar parameters make possible the study of the metallicity distribution of {\it Kepler} targets \cite{dong14} and the eccentricity distribution of {\it Kepler} planets \cite{xie16}. Stellar parameters have also been extracted from LAMOST spectra using pipelines other than the official LASP, such as LSP3 \cite{lsp3} and ROTFIT \cite{rotfit}. Using ROTFIT parameters, \cite{rotfit2} found that on average the hosts for hot Earth-size planets have super-solar metallicity.

We study host metallicity distribution for {\it Kepler} planets in the period range of $1 {\,\rm d} < P < 10 {\,\rm d}$ using the LAMOST/LASP stellar parameters \cite{luo} from Data Release 4 (DR4) \footnote{http://dr4.lamost.org}, which contains data for $\sim 60,000$ main-sequence {\it Kepler} stars with spectroscopic types AFGK ($\sim30\%$ of of all {\it Kepler} targets). Previous studies \cite{BN13, Maz16} have identified a ``desert'' of short-period Neptunes ($P = 2-4 \,{\rm d}$ and it may extend to $5-10 \,{\rm d}$) by studying samples from Radial-Velocity (RV)  surveys, mixed ground-based and {\it Kepler} transit searches. We benefit from the homogeneity of the {\it Kepler}  sample with well quantified detection efficiencies as well as accurate stellar and planetary parameters thanks to LAMOST, and this work is focused on the effects of metallicity.

We discover a  ``cousin'' population of hot Jupiters: like hot Jupiters, these short-period Neptune-size planets ($2 R_\earth \lesssim R_p \lesssim 6 R_\earth$), dubbed as ``Hoptunes'', are predominantly hosted by metal-rich stars, and they also reside more often in single transiting than multiple transiting planetary systems. In the radius distribution the populations of hot Jupiters and Hoptunes frame a ``valley'' at approximately Saturn size (in the range of $6 R_\earth \lesssim R_p \lesssim 10 R_\earth$), and in this ``hot-Saturn valley'' there is a significant deficit in planet frequency. 

\section*{Sample Selection}

We select our stellar sample using the official LASP \cite{luo} parameters from LAMOST DR4 AFGK stellar catalog. The stars satisfy $4700 K < T_{\rm eff} < 6500 K$ and $\log g > 4.0$ so that our sample consists primarily of solar-type main-sequence stars. They also cover a range of stellar parameters for which we have comparison stars available with accurate high-resolution spectroscopic parameters \cite{dong14, xie16}. Our stellar sample consists of 30,727 stars in total. Stellar masses and radii are derived by isochrone fitting as described in \cite{xie16}. We then cross-match with the catalog of {\it Kepler} planet candidates and remove the known false positives and those with large false positive probabilities (FPPs). Stellar parameters derived from LAMOST/LASP are applied to derive the planet properties such as planet radius $R_p$, which has a typical error of about $15\%$. We focus on those with orbital period $1 {\,\rm d} <P< 10 {\,\rm d}$ and planet radius $1 R_\earth <R_p< 20 R_\earth$. The sample has 295 planets in 256 systems. 151 planets are in systems that contain single {\it Kepler} transiting planets, and 144 planets are in multiple transiting planet systems where the other transiting planets in the systems are not necessarily in the above-mentioned period and radius ranges. In the Section 1 of the Supporting Information (SI), we provide more details on sample selection, and we also show that the main results discussed below are not sensitive to the exact choices of sample selection. The LAMOST stellar parameter and planet parameters of our sample are given in Table S1 of SI. 


\section*{Analysis \& Results}
We divide the stellar sample into two metallicity subsamples: metal rich ([Fe/H]$\ge$0) and metal poor ([Fe/H]$<$0). Figure 1 shows the distributions of period and planet radius for the planets that belong to the two subsamples, respectively. 

\begin{figure}[tbhp!]
\includegraphics[width=1.0\linewidth]{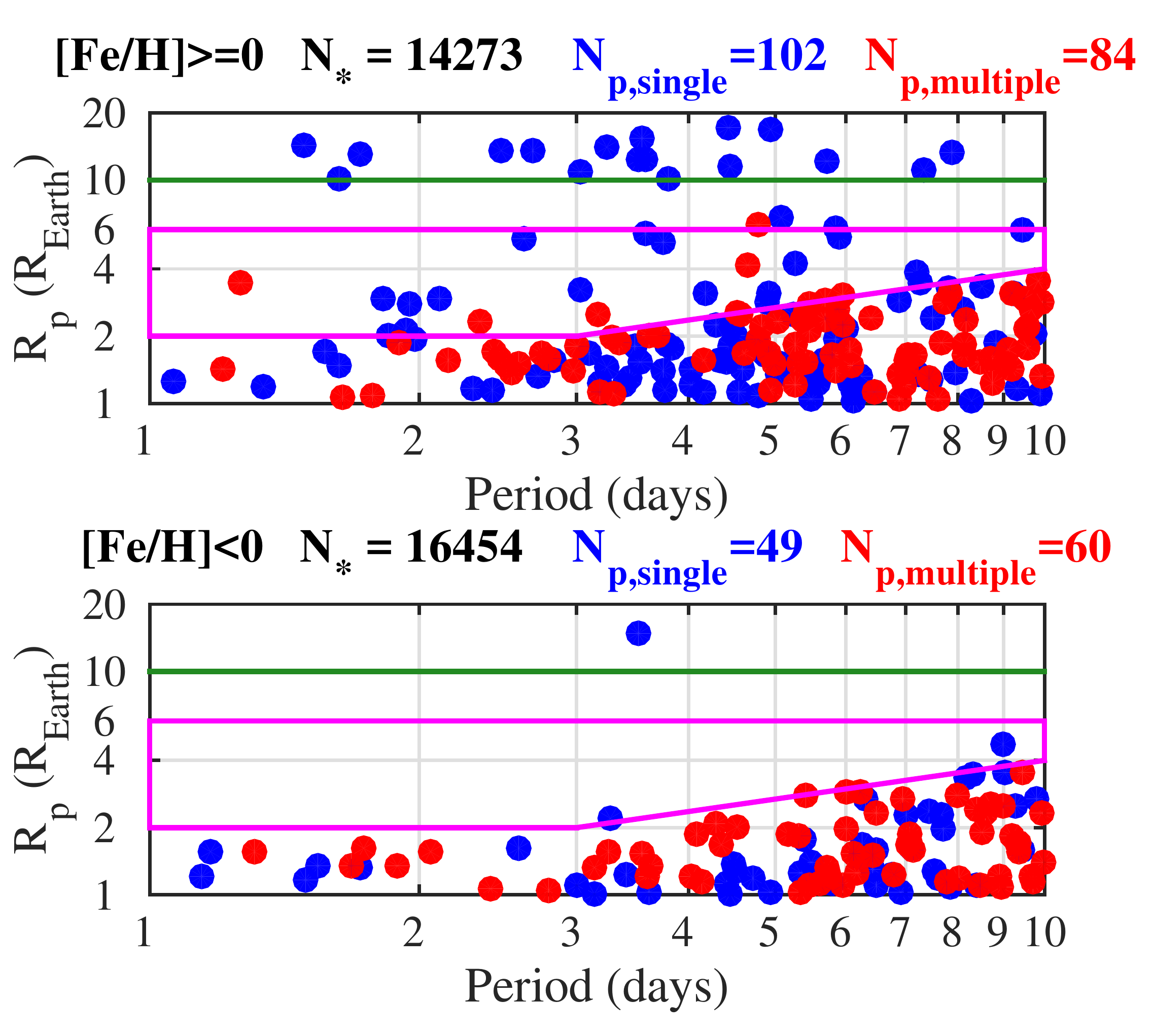}
\caption{Period-radius distribution for short-period {\it Kepler} planet candidates hosted by metal-rich (top)  and metal-poor (bottom) stars. Planets in single and multiple transiting planetary systems are plotted in blue and red circles, respectively. The numbers of planets and stars (including non-transiting targets) are printed on the top of each panel.
The dark green horizontal line ($R_p=10R_\oplus$) denotes the {\it empirical} lower boundary of hot Jupiter, and the magenta lines denote the empirical boundaries of Hoptunes (see Section 3). Note that the typical uncertainty in $R_p$ is about $15\%$.}
\end{figure}

In the metal-poor subsample, except for a handful of objects, almost all the planets lie in the bottom part of the period-radius plane: for those with $1 {\,\rm d} <P< 3 {\,\rm d}$, their radii are smaller than about $\approx 2\,R_\earth$, and for those with $3 {\,\rm d} <P< 10 {\,\rm d}$, the upper boundary in radius grows from $\approx 2 R_\earth$ to $\approx 4\,R_\earth$ with increasing period. This boundary is approximately illustrated as a magenta polyline connecting $(P, R_p)$ = $(1,2)$ with $(3,2)$ with $(10,4)$ on the period-radius plane. At $P>3\,$d, this boundary is similar to the lower boundary of the ``desert'' described in \cite{Maz16} (see Figure S4 in SI). In the metal-rich subsample, about 3/4 of the planets are concentrated below the magenta polyline. The rest $\sim 1/4$ of all planets in the metal-rich subsample reside in the parameter space that is sparsely   populated with planets for the metal-poor subsample. One population of planets hosted by these metal-rich stars have radii larger than $\sim 10\,R_\earth$ (marked by a dark green solid line), and they are the hot Jupiters. The other population hosted by metal-rich stars have radii smaller than $\sim 6\,R_\earth$ while larger than those indicated by the bottom magenta polyline. Their sizes $(2-6\,R_\earth)$ are close to that of the Neptune at about $4\,R_\earth$,  but we do not know whether all of them are physically Neptune-like planets. Given the uncertainties in their physical states (Neptunes, Super-Earths, mini-Neptunes, or other possibilities), we choose to dub this population as ``Hoptunes'' in the text. This name is to reflect our current level of understanding of this population -- without mass measurements, we do not know whether they are physically Neptunes, and so far, they can be only isolated in a specific regime in the period-radius plane as found in this study.

Next we discuss three main features in the distributions, including two striking similarities between Hoptunes and hot Jupiters, and a significant ``valley'' separating them in the radius distribution.

{\bf 1) Hoptunes and hot Jupiters are both preferentially hosted by metal-rich stars, and for both populations the planet frequencies correlate with the host [Fe/H].}

The contrast between metal-rich and metal-poor stars in hosting hot Jupiters and Hoptunes is statistically significant. In the metal-rich subsample, there are 17 hot Jupiters and 24 Hoptunes, which are $9.1^{+2.8}_{-2.2}\%$ and $12.9^{+3.2}_{-2.6}\%$ of all planets in that subsample, respectively, with the uncertainties inferred from the Poisson distribution.  If the metal-poor subsample had similar relative fractions of these two populations, we would expect to see $10.0^{+3.0}_{-2.4}$ hot Jupiters and $14.1^{+3.5}_{-2.9}$ Hoptunes, but there are only one hot Jupiter and two Hoptunes detected. Figure 2 shows the cumulative fraction of various planet hosts and the stellar sample as a function of [Fe/H]. The strong dependence of hot Jupiters' occurrence on host-star metallicity is well known, and such a dependence is clearly seen from our sample. As compared with the stellar sample as well as ``other hot planets'' (i.e., not hot Jupiters or Hoptunes), the hot Jupiters are preferentially hosted by metal-rich stars. Using the two-sample Kolmogorov-Smirnov (K-S) tests, the [Fe/H] distributions of hot Jupiters are inconsistent with drawing from the same sample for the stellar sample and ``other hot planets'' (at p-values of $3\times10^{-5}$ and $3\times10^{-4}$, respectively). Hoptunes have similarly strong preference for metal-rich hosts, and their distributions are inconsistent with the stellar sample and ``other hot planets'' with even higher statistical significance (at p-values of $4\times10^{-4}$ and $3\times10^{-4}$, respectively). The two sample K-S test on the cumulative distributions of hot Jupiters and Hoptunes results a p-value $45\%$ so we cannot reject the null hypothesis that the host metallicities of Hot Jupiters and Hoptunes are drawn from the same distribution.
The distribution of ``other hot planets'', dominated by Earth-size planets ($1 R_\earth <R_p <2 R_\earth$), is also more skewed (though in a degree significantly less than hot Jupiters and Hoptunes) toward higher metallicities as compared to the stellar sample (the K-S test p-value $=0.002$), which is qualitatively consistent with the conclusion by \cite{rotfit2} that there is a preference of metal-rich hosts for hot Earths.
Later in this section we derive the intrinsic frequencies by taking into account the incompleteness of {\it Kepler}.

\begin{figure}[tbhp!]
\includegraphics[width=1.0\linewidth]{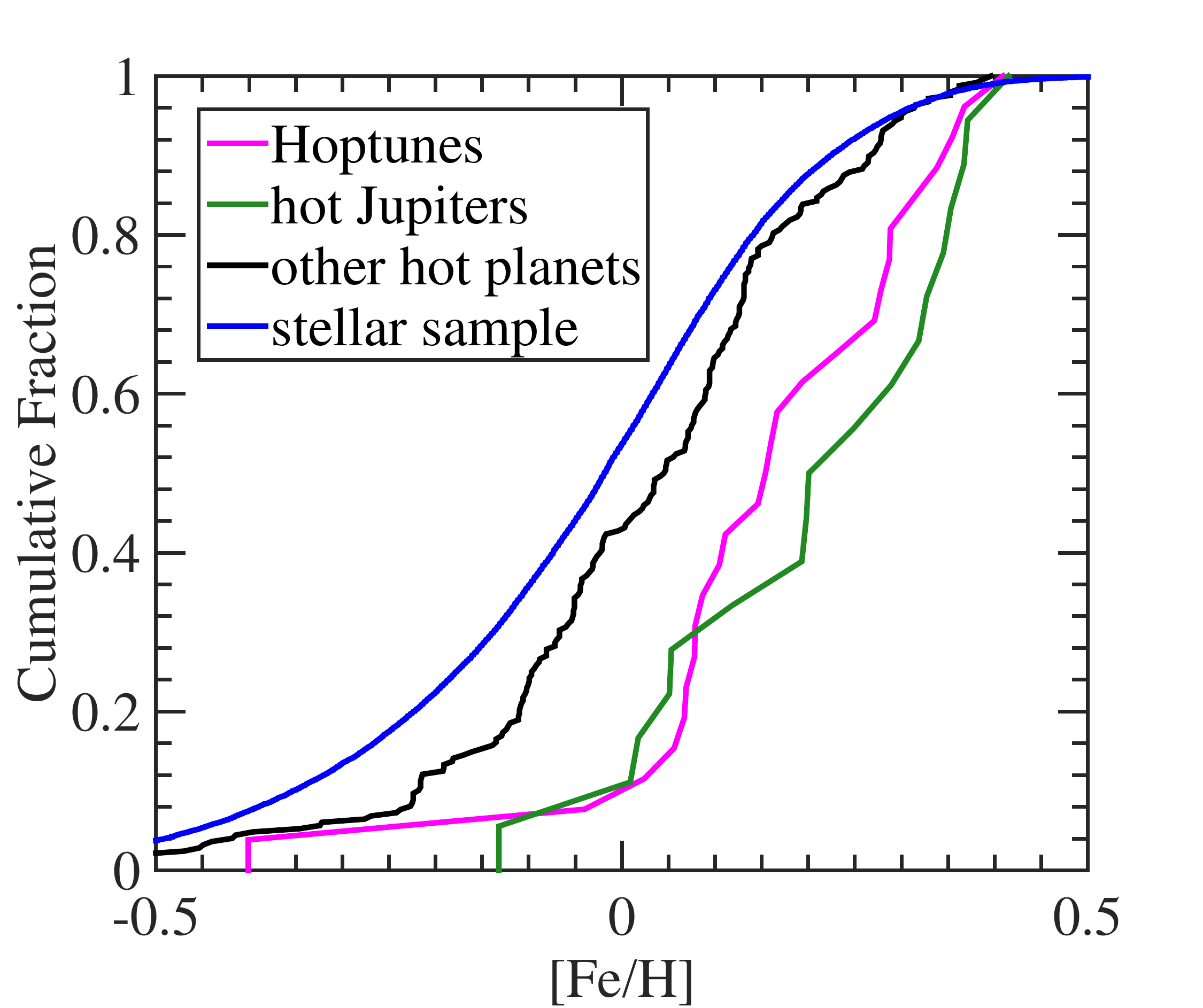}
\caption{Cumulative fractions as a function of host [Fe/H]. Hoptunes (magenta) and hot Jupiters (dark green) have similar distributions, and using the two-sample K-S test, the two samples are drawn from the same underlying
distribution with a probability of $45\%$. In contrast, the distributions for both Hoptunes and hot Jupiters are different from those of the stellar sample (blue) and other planets (black) at high statistical significances.}
\end{figure}

{{\bf 2) Hoptunes and hot Jupiters both tend to preferentially exist in {\it Kepler}'s single transiting planetary systems.} }

A distinguishing feature of hot Jupiters is that they tend not to have neighboring planets in close-by orbits \cite{lathamhotj,lonely,warmhot}. All the hot Jupiters in our sample are single transiting planet systems, and the majority of the Hoptunes are in {\it Kepler} singles ($73 \pm 9\%$ with the uncertainty estimated from binomial distribution). In contrast, in the regimes under the magenta polylines, slightly less than half of planets ($45\pm 3\%$) are in single transiting planetary systems -- note that the single fraction are similar for both metal-poor and metal-rich subsamples, approximately $46 \pm4 \%$ for metal-poor and $43 \pm 5\%$ for metal-rich. Therefore, like hot Jupiters, Hoptunes are also preferentially in singles compared to other hot planets. The single fraction of Hoptunes is smaller than that of hot Jupiters in our sample. We employ a likelihood analysis based on binomial distribution to test the statistical significance of this difference, and the null hypothesis that they have the same single fraction is ruled out at $2.9\,\sigma$ significance (see the Section 2 of the SI for detail).

{\bf 3) Hoptunes and hot Jupiters are separated by a ``hot-Saturn valley". } 

As can be seen in Figure 1, there is a deficit of planets near Saturn size (about 6--10\,$R_{\earth}$) in the planet-radius distributions between the Hoptunes and hot Jupiters, and we refer this deficit as the ``hot-Saturn valley". The hot-Saturn valley can be clearly seen in the cumulative distribution of planet radius for planets with $R_p > 4\,R_{\earth}$ from our sample (upper panel of Figure 3). The cumulative distribution has two clear breaks occurring at $\sim 6.5\,R_{\earth}$ and  $\sim 10\,R_{\earth}$, and between these breaks, planets occur significantly less frequently compared to other ranges of radii. In the middle panel of Figure 3, we show a histogram of observed radius distribution with 12 uniformly spaced bins in logarithm. The four bins closest to $8\,R_{\earth}$ have one planets where the four bins on the left (Hoptunes) have 10 planets and the four bins on the right (hot Jupiters) have 16 planets. If the significance of the valley is assessed by the difference between adjacent sets of bins, we can adopt the Skellam distribution to calculate the probability assuming a null hypothesis of all bins having the same expected numbers of planets. Comparing 10 or 16 planets in the neighboring sets of bins with 1 planet in the central bins, we find that the probability of the valley due to random fluctuations is about $1\%$. The existence of such a valley with relatively sharp boundaries also suggests that the (unknown) effects of blending due to un-resolve background binaries are unlikely significant in blurring the planet radius distribution.

\begin{figure}[tbhp!]
\includegraphics[width=1.0\linewidth]{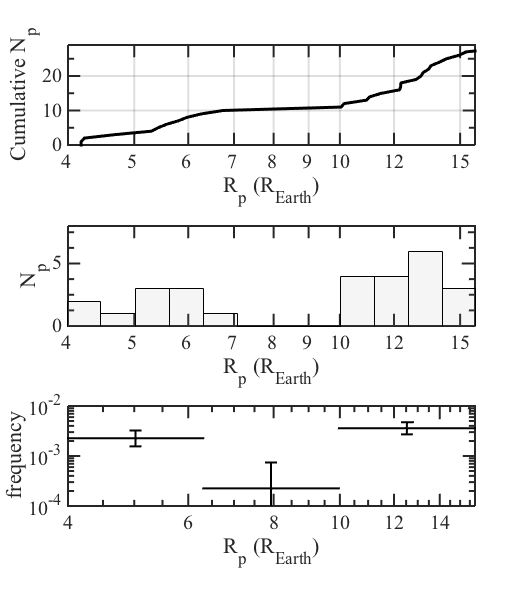}
\caption{The hot-Saturn valley revealed from the radius distribution of planets in our sample.  Top: Cumulative distribution of planets with radii larger than 4 $R_\oplus$.  Middle: Number of detected planets in radius bins with equal size in logarithm. Bottom: Intrinsic planet frequencies as a function of planetary radius.}
\end{figure}

Finally we consider the incompleteness of {\it Kepler} survey and calculate the intrinsic planet frequency for our sample. We take the survey selection, which is the incompleteness due to the survey detection thresholds, and the  transit geometric bias into account. In order to calculate the incompleteness, we use the code supplied by \cite{Bur15} based on the detection efficiency characterization method of \cite{chris15} to calculate the correction factor per bin in the parameter space, and we apply the LAMOST stellar parameters in the calculations.

\begin{figure}[tbhp!]
\includegraphics[width=1.0\linewidth]{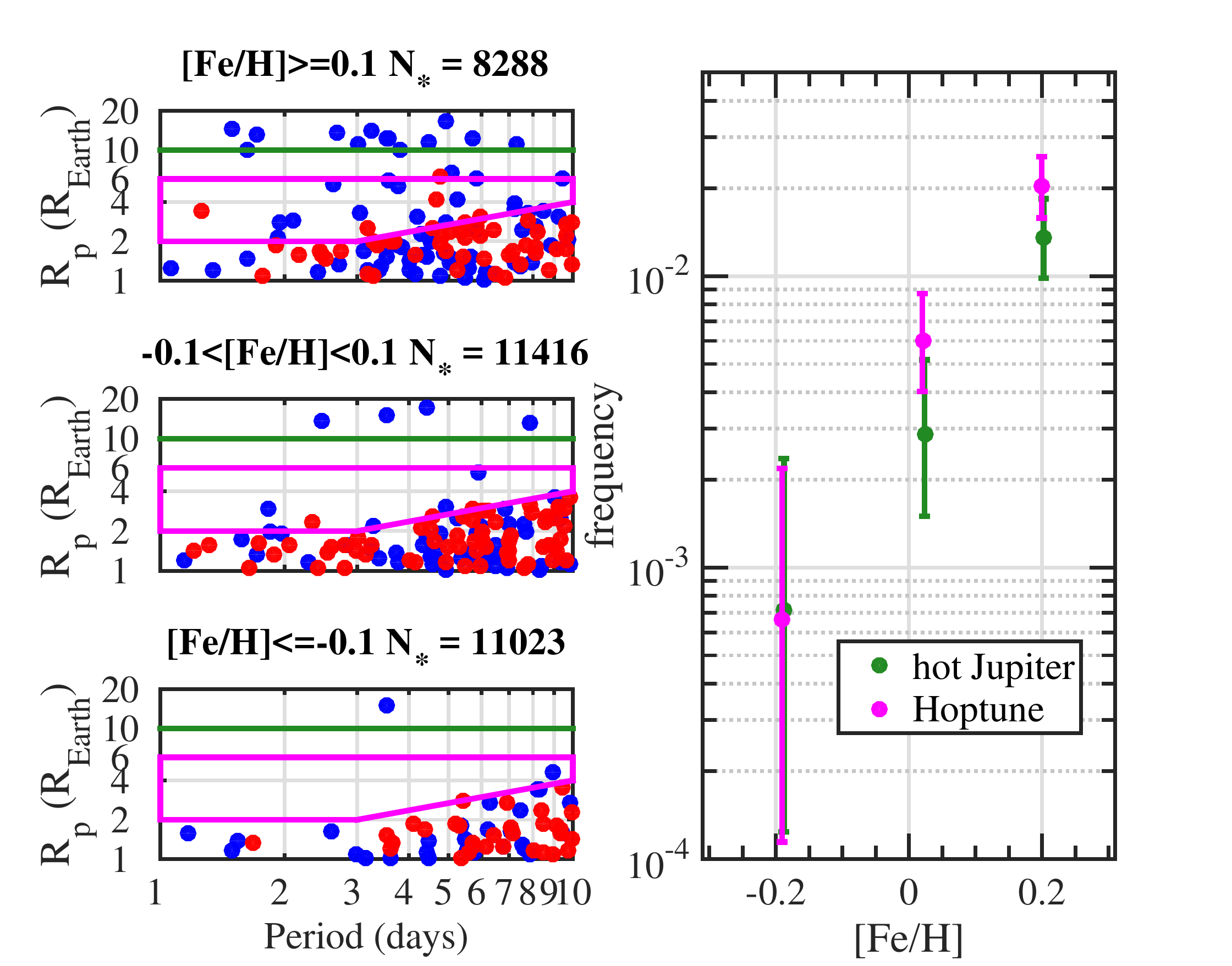}
\caption{The dependence of planet distribution (Left Panel) and intrinsic frequency (Right Panel) on stellar metallicity ([Fe/H]) using three metallicity subsamples. The frequencies of hot Jupiters and Hoptunes have similar trends as a function of metallicity -- for both populations, they increase by a factor $\sim 10$ from the ``sub-solar'' metallicity regime ([Fe/H]$\leq-0.1$) to the super-solar ([Fe/H]$\geq0.1$) regime, and the frequencies of Hoptune are similar (within a factor of $\sim 2$) to those of hot Jupiters for all subsamples.} 
\end{figure}

To better examine the dependence of the planet frequency on metallicity, we divide the sample according to [Fe/H] into three subsamples -- ``super solar'' ([Fe/H]$\ge$0.1), ``solar'' ($-0.1<$[Fe/H]$<0.1$) and ``sub solar'' ([Fe/H]$\le-0.1$). The planet period-radius distributions for three metallicity subsamples are shown in the left three panels of Figure 4. In the ``sub-solar'' subsample (the bottom-left panel), there are only a handful of hot Jupiters and Hoptunes. Then in the ``solar'' subsample (the middle-left panel), there are several hot Jupiters, and a few Hoptunes start to emerge close to the lower boundary (the magenta polyline). Finally, in the ``super-solar'' subsample (the upper-left panel), there is a dramatic increase in the numbers for both hot Jupiters and Hoptunes. A noticeable trend is that the lower boundary of planet ``valley'' evolves from $\approx 6 R_{\earth}$ (at around Saturn size) in the ``super-solar'' subsample to lower values, close to $\approx 3-4 R_{\earth}$ (at around Neptune size) in the ``solar'' subsample. The right panel shows the intrinsic frequencies of  hot Jupiters and Hoptunes after correcting the survey incompleteness and geometric biases. The frequencies of Hoptunes and hot Jupiters increase dramatically (by a factor of $\sim10$ from ``sub-solar'' to ``super-solar'') with [Fe/H]. Such a trend for hot Jupiter hosts is well known (see e.g.  \cite{guojup} and the references therein), and the trend for Hoptune hosts is remarkably similar. The frequencies of Hoptunes are similar (within a factor of $\sim 2$) to that of hot Jupiters for all subsample. Hosts of Hoptunes and hot Jupiters  have similarly strong preferences for host metallicities higher than the sun (``super-solar'' vs. ``solar'').

The bottom panel of Figure 3 shows the intrinsic frequencies of planets within the $\log(R_p)$ bins spanning $0.2$ dex around $\sim5\,R_{\earth}, \sim8\,R_{\earth}$ and $\sim12\,R_{\earth}$ in our sample. The averaged frequencies of the planets in these subsamples are $0.23^{+0.10}_{-0.07}\%$, $0.02^{+0.05}_{-0.02}\%$, $0.36^{+0.10}_{-0.09}\%$ at approximately Neptune size, Saturn size and Jupiter size, respectively. Thus the hot-Saturn valley represents an approximately one order of magnitude depression in planet frequency as a function of planet radius. The averaged planet frequencies ${\rm d} N_p/{\rm d} \log(R_p) {\rm d} \log(P)$ for hot Jupiters and Hoptunes in these subsamples are consistent within the uncertainties. 

\section*{Discussion \& Conclusion}

Our findings benefit from the homogeneity of the {\it Kepler} sample as well as accurate planet radii and host metallicities thanks to the high accuracy of LAMOST stellar parameters. \cite{BN13} and \cite{Maz16} found  a desert 
of short-period Neptunian planets using mixed samples of planets discovered from various surveys. The upper boundary of the desert defined by \cite{Maz16} is similar to that of the hot-Saturn valley we identify in this work (see Figure S4 in SI). While this upper boundary from \cite{Maz16} is a diagonal line on the period-radius plane based on a large number of planets mostly contributed from ground-based transits, our sample is too small to clearly determine a slope. Their lower boundary resembles the lower boundary of Hoptunes, especially for $P>3 \,{\rm d}$. 
According to our results, their desert appears to encompass both the hot-Saturn valley and the Hoptunes. Inside their desert, we find that Hoptunes have a similar averaged frequency of $\sim 1\%$ as hot Jupiters. In comparison, hot-Saturn valley has an order-of-magnitude deficit in planet frequency. These features appear to be largely ``washed out''  in \cite{Maz16}. This is mainly due to the large uncertainties and systematics biases in $\log$(g) measurements of the KIC catalog (e.g., see Fig. S4 of \cite{xie16}), and the resulting large errors in planet radii fill some objects in the valley (see the right panels of Fig. S3 in SI).

The similarities in host metallicity, intrinsic frequency and preference for single transiting planetary systems suggest a close link between hot Jupiters and Hoptunes in their migration and/or formation processes.

The correlation between the intrinsic frequency of short-period Jupiter and stellar metallicity has been well established \cite{santos04, fischer05}. 
According to core-accretion models, the total masses of building-block planetesimals and/or embryos are proportional to those of the heavy elements in the host stars \cite{IL04b}. One possible interpretation of the metallicity correlation for short-period Jupiters is that the metal-rich environment provides more building blocks to form massive planetary cores ($\sim$10 $M_\oplus$) before the gas disk dissipates, which are crucial for gas accretion to form giant planets like hot Jupiters \cite{Pol96}.  
However, such an interpretation may not be well suitable for the metallicity correlation of Hoptunes, since for most Hoptunes, especially those smaller ones with radius less than 4$R_\earth$, they do not need massive cores to accrete as much gas as needed for forming Jupiters. 
Another possibility is that metallicity may play an important role to trigger/amplify certain migration mechanisms for hot Jupiters \cite{DMC13, Liu16}. Such mechanisms should then similarly operate for Hoptunes and also preferentially produce single transiting planetary systems. Note that we find at $2.9 \sigma$ significance that the fraction of single-transiting planet systems of hot Jupiters is higher than that of Hoptunes, indicating that whatever process removes the ``brothers'' of hot Jupiters likely operate less efficiently for Hoptunes.
The {\it in situ} formation mechanisms (e.g., \cite{insitun}) are also subject to these constraints too.

The radius range of the hot-Saturn valley ($R_p$$\sim$6--10$R_\oplus$) roughly corresponds to the mass domain (between $\sim 10-30$ and $\sim 100-200$ $M_\oplus$) of the ``planet desert" expected from core accretion simulations of planet formation  \cite{IL04a, Mor09}. However, these predictions apply to $a=0.2{\rm} - 3 \,{\rm AU}$ while the planets in our sample have $a<0.1 \,{\rm AU}$. We plan to investigate whether the hot-Saturn valley extends to longer period in a future work.

When studying the radius distribution for short-period planets, it is important to consider the effects of planetary inflation (e.g., \cite{LF16} and the references therein) and/or evaporations \cite{evap, OW13, Jin14}.  \cite{evap} suggest that hot Neptunes may originate from evaporation of hot Jupiters, and thus they may share common origins and evolution history (though note that \cite{evap} has been contradicted by some follow-up theoretical studies such as \cite{ruth}). This hypothesis is consistent with the similarities between hot Jupiters and Hoptunes found in this work. In addition, it may be interesting to test whether mechanisms such as photo-evaporation can be responsible in shaping features such as the sharp lower boundary for Hoptunes (the magenta polyline in Figure 1) in the period-radius distribution and also explain how the sharpness of this boundary varies with host metallicity. For instance, the consequence of photo-evaporation may depend on planet core mass \cite{OW13, Jin14}, and one may speculate that core mass distribution can differ according to host [Fe/H] thus can potentially play a role in forming the [Fe/H]-dependent planet distribution found here. RV follow-ups of Hoptunes can provide the crucial mass measurements to reveal their physical states and test such scenarios. 

The ``kinship'' between hot Jupiters and Hoptunes as well as the hot-Saturn valley separating these two cousins offer unique clues and constraints for the formation and migration of short-period planets. Future surveys and missions, particularly Transiting Exoplanet Survey Satellite (TESS), 
are expected to detect many more short-period planets and explore these features with greater details.

\textbf{Acknowledgements}
We thank A. Gould for stimulating discussions and insights into statistics. We thank the three anonymous referees for their helpful reviews. We are grateful to Dan Huber,  Josh Winn, Yanqin Wu, Dong Lai, Wei Zhu, B. Katz, K. Stanek, S. Kozlowski, C. Mordasini and D. Lin for helpful comments. S.D. acknowledges Project 11573003 supported by NSFC and the LAMOST Fellowship, which is supported by Special Funding for Advanced Users, budgeted and administrated by Center for Astronomical Mega-Science, Chinese Academy of Sciences (CAS).  J.-W.X. and J.-L.Z. acknowledge support from the Key Development Program of Basic Research of China (973 Program, Grant 2013CB834900) and the NSFC Grants 11333002. J.-W.X. is also supported by the NSFC Grant 11403012 and a Foundation for the Author of National Excellent Doctoral Dissertation of People's Republic of China. Guoshoujing Telescope (the Large Sky Area Multi-Object Fiber Spectroscopic Telescope; LAMOST) is a National Major Scientific Project built by the CAS. Funding for the project has been provided by the National Development and Reform Commission. LAMOST is operated and managed by the National Astronomical Observatories, CAS.



\bibliography{hoptune.bib}

\cleardoublepage

\beginSI
\begin{center}
{ \LARGE  Supporting Information (SI)}\\[0.5cm]
\end{center}

\section{On the Selection of Stellar and Planet Sample}

We cross-match between the AFGK Stars Catalog of LAMOST DR4 and all target stars observed during Q1-Q16 of the {\it Kepler} mission. As a result, there are 59,964 unique {\it Kepler} target stars with LAMOST LASP stellar parameters. We pick stars satisfying $4700 K < T_{\rm eff} < 6500 K$ and $\log g > 4.0$ so that the stellar sample is mostly main-sequence stars, and 30,727 stars survive the cut (green points the A1 panel of Figure S1). We also test other selection criteria of the stellar samples to check if they affect our main results. In one test, we choose $\log g > 4.3$ to more aggressively remove sub giants in the sample (green points the A2 panel of Figure S1). In another test, we use a sloped upper boundary in $T_{\rm eff}$-$\log g$ (green points in the A3 panel). The resulting planet distributions for [Fe/H]$>=0$ and [Fe/H]$<0$ are plotted in B2) and C2), B3) and C3) panels, respectively. The main results of the paper remain the same and are not sensitive to the specific choice of the stellar sample. 

We cross-match the LAMOST AFGK catalog with the Kepler planet-candidate catalog.  We primarily use the planet sample from Q1-Q16  \cite{Mul15}, which has well quantified detection efficiencies \cite{chris15, Bur15}. The following updated catalog DR 24 \cite{Cou16} is flawed for occurrence rate calculations \cite{chrisdr24}. The most recent catalog DR 25 \cite{Twi16} is not yet finalized at the time of our work. We have nevertheless used DR24 and DR25 catalogs to make consistency check, and we find excellent agreements with our results using Q1-Q16. There are 2422 Kepler Object of Interests (KOIs) in 1996 systems with LAMOST stellar parameters. Then we apply a series of selections. 
First of all, we attempt to remove false positives (FPs). We remove KOIs if they are identified as FPs according any of the following references: the Q1-Q16 catalog \cite{Mul15}, the DR 24 catalog \cite{Cou16}, the DR 25 catalogs \cite{Twi16}, and the SOPHIE RV follow-up survey of giant planets \cite{San16}. For the planet sample analyzed in the main text, we also follow \cite{xie16} and remove the candidates with large False Positive Probabilities FPP$>68\%$ in the Kepler Astrophysical False Positive Probabilities Table \cite{Mor16}. We note that eliminating candidates with too small FPP may potentially bias planet statistics by significantly removing a large fraction of true planets (e.g., FPP$=10\%$ means that $90\%$ probability being true planets). Nevertheless, we also test removal of candidates with FPP$>1\%$ while keeping all other criteria unchanged. As shown in Figure S2, the resulting distribution (right two panels) hardly differ from that of the default planet sample (left two panels). We also remove KOIs with transit signal-to-noise ratios (SNRs) smaller than 7.1 and radius $R_p>20 R_\oplus$. After these cuts, we are left with 1218 KOIs in 913 systems.  Next, we apply the same stellar parameter cut as the stellar sample ($4700 K <T_{\rm eff}< 6500 K$ and $\log g>4.0$), and then we have 945 KOIs in 686 systems. Finally, we only analyze planets with periods between 1 and 10 days and radii greater than 1 $R_\earth$. The reason for excluding planets with periods shorter than $1\,{\rm d}$ is because the standard Kepler pipeline is not well conditioned to detect these ultra short period (USP) KOIs \cite{Jac13, SO14, Ada16}. Finally, we obtain a sample with 295 planets in 256 systems.

As discussed in the main text, accurate LAMOST stellar parameters are the key to carry out this study. Figure S3 shows the distributions of the planet candidates in our sample when KIC stellar parameters are used. Due to the large uncertainties and systematic errors in $\log g$ and [Fe/H], it is impossible to identify the patterns in planet distributions discovered in this work.

The planet detection efficiency is calculated using the Python code (\hyperlink{}{https://github.com/christopherburke/KeplerPORTs}) by \cite{Bur15} with stellar parameters replaced by those of the LAMOST. $R_*$ and $M_*$ are derived following the methods described in \cite{xie16}. It is worth noting that the survey efficiencies are essentially the same in the regimes of interests for metal-rich and metal-poor systems (Figure S4). 

\section{On the statistical significance of  a larger single-planet fraction of hot Jupiters than that of Hoptunes}

In our sample, there are 18 hot Jupiters, and all of them are in single transiting-planet systems. In comparison, there are 26 Hoptunes in total, and 19 of them are in single transiting-planet systems. Below we construct a likelihood analysis based on binomial distributions to assess whether the single-planet fraction of Hoptunes being  higher than that of hot Jupiters is statistically significant.

We denote the number of single-transiting and multiple Hoptunes as $m_1$ and $m_2$, respectively and the number of single-transiting and multiple hot Jupiters as $n_1$ and $n_2$, respectively. 

For the two sets of binomial measurements with $(m_1, m_2)$ in one and $(n_1, n_2)$ in the other, if we fit the single fractions separately, then the best estimate of single fractions are $p_m = m_1/M$ (where $M=m1+m2$) and $p_n = n_1/N$ (where $N=n_1+n_2$). If we fit them jointly and to find the best-fit overall single fraction $p$, then likelihood based on binomial distribution is $L = C^{m_1}_M {m_1}^p {m_2}^q C^{n_1}_N {n_1}^p {n_2}^q$, where  $(q\equiv 1-p)$. 

Setting $d\ln L/dp = 0$, one can easily find the best-fit $p = (m1+n1)/(M+N)$. Then the likelihood ratio for introducing an extra parameter, that is $p_n$ and $p_m$ instead of just $p$ is $L_2/L_1 = (p_m/p)^{m_1}\times (q_m/q)^{m_2} \times (p_n/p)^{n_1} \times (q_n/q)^{n_2}$ 

In the present case, $n_2$ = 0 and $N=n_1$, therefore,  $\ln(L_2/L_1) = \ln(p_m/p)^{m_1} \times (q_m/q)^{m_2} \times (p_n/p)^N) = \ln A + \ln B$ where, $\ln A = m_1\times \ln ((m_1\times(M+N))/(M\times(m_1+N)))+ m_2\times \ln((M+N)/M)$ and $\ln B = N\times \ln ((M+N)/(m_1+N))$.

Plugging the numbers ($m_1=19$, $m_2=7$, $M=26$, $N=n_1=18$), we obtain a logarithm likelihood ratio $\ln(L_2/L_1)=4.1$, corresponding to about $2.9\sigma$ significance.

\begin{figure*}[t]
\centering
\includegraphics[scale=0.7]{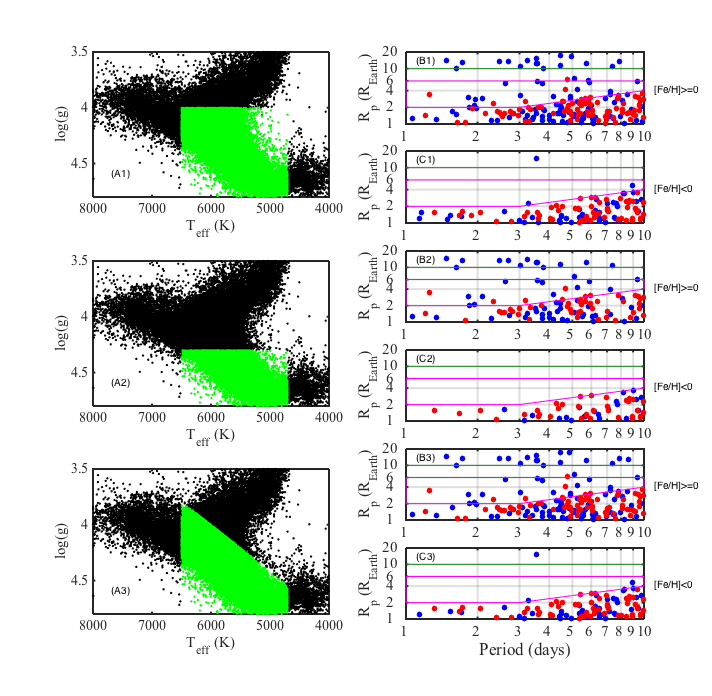}
\caption{Stellar sample selections and tests. The green dots in Panel (A1) show the star selected in the main text, and they satisfy $4700 K < T_{\rm eff} < 6500 K$ and $\log g > 4.0$. We test the effects of alternative selections. The green dots in Panel (A2) show a more stringent criterion of $\log g > 4.3$ to more aggressively remove sub giants. The resulting distributions in planets (B2 and C2) are similar to those from the default sample shown in panels B1 and C1. The results (shown in B3 and C3) are similar for sloped upper boundary in $T_{\rm eff}$ and $\log g$ plane (A3). }
\end{figure*}

\begin{figure*}[t]
\centering
\includegraphics[scale=0.45]{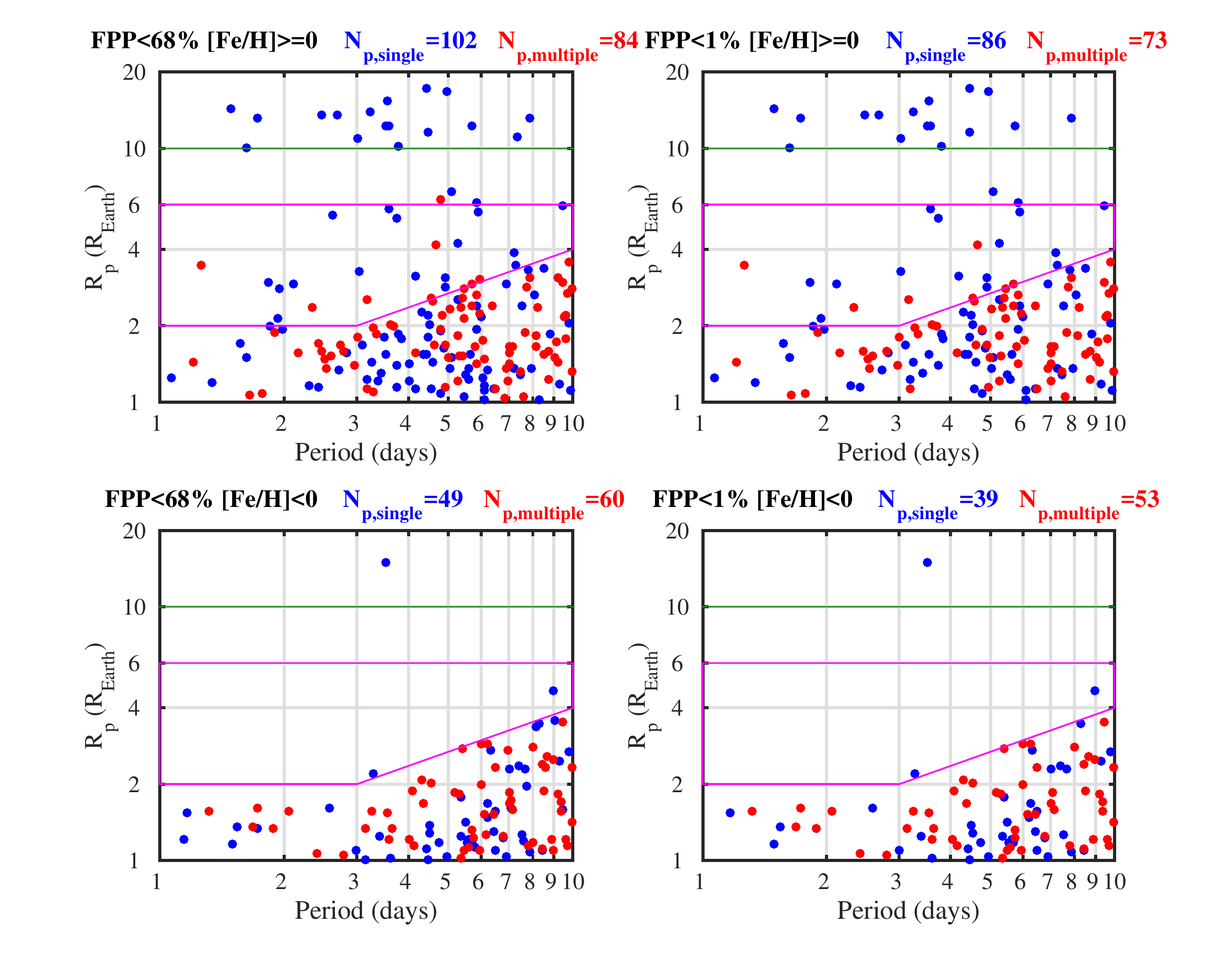}
\caption{The right panels show the resulting planet distributions for a more stringent cut for False Positive Probability FPP$<1\%$, and the result distributions are similar in main features with the default selection criteria FPP$<68\%$.}
\end{figure*}

\begin{figure*}[t]
\centering
\includegraphics[scale=0.45]{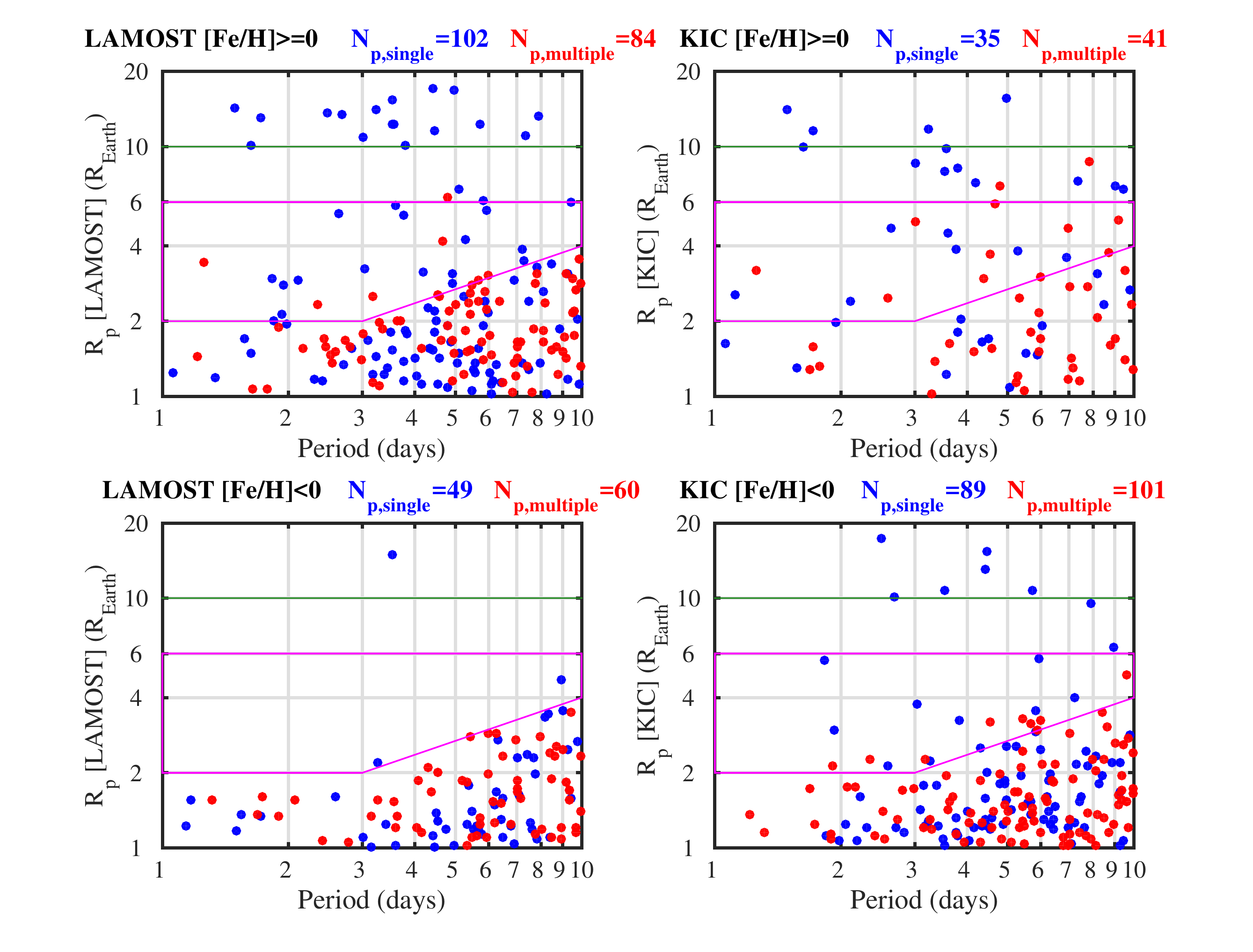}
\caption{The right panels show the planet distributions when the stellar parameters from the KIC catalog are adopted. Due to the large uncertainties and systematics in the KIC parameters, it is impossible to identify the main patterns in the planet distributions concluded in this work using the accurate LAMOST parameters (the left panels).}
\end{figure*}

\begin{figure*}[t]
\centering
\includegraphics[scale=0.55]{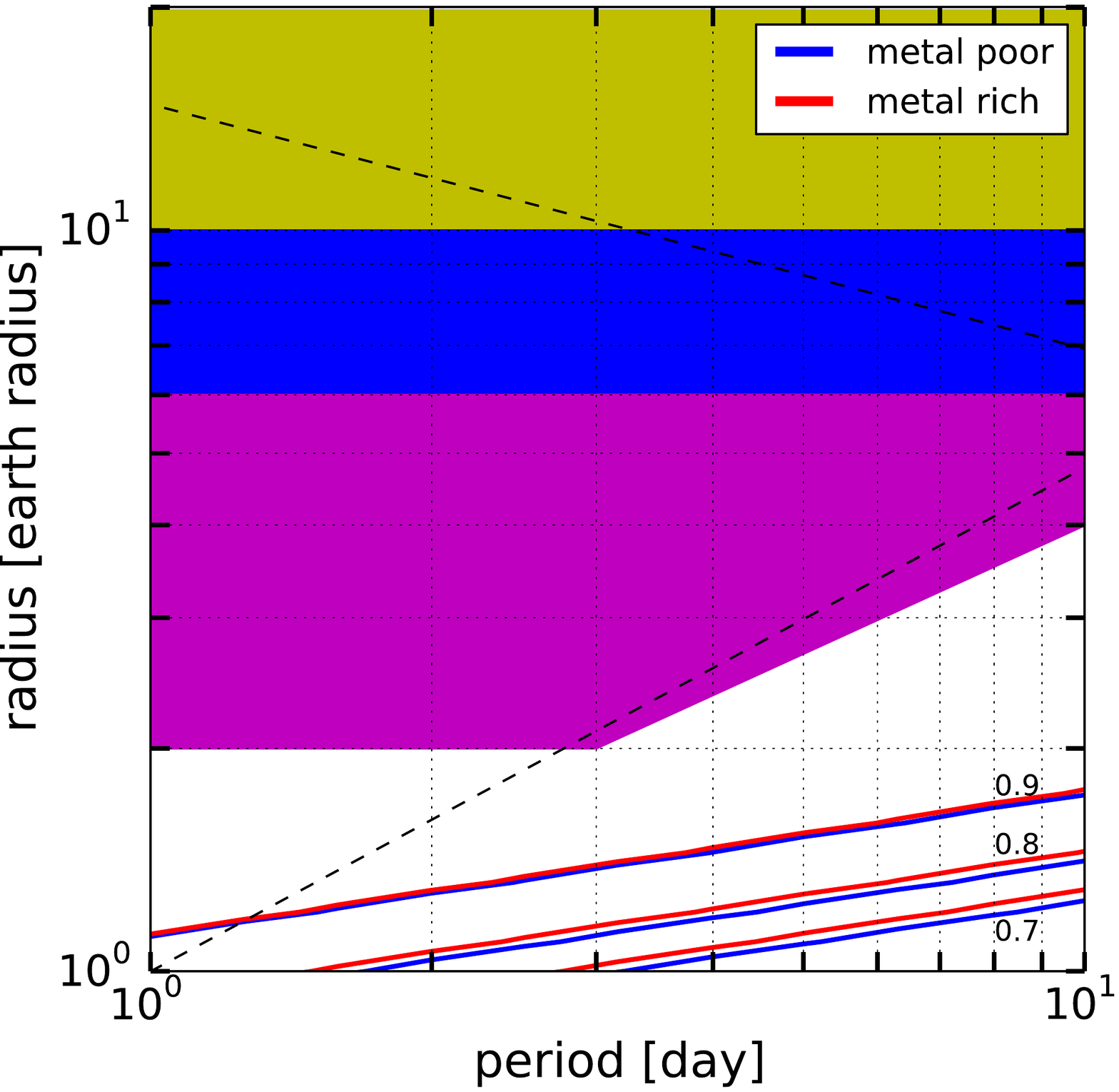}
\caption{The Kepler survey efficiency for the metal poor ([Fe/H]$<$0) and metal rich ([Fe/H]$\ge$0) subsamples are shown as contours in blue and red, respectively, in the period-radius diagram. The efficiencies are nearly identical for the two subsamples in the regimes of our interest. The regimes of Hoptunes, hot Saturns and hot Jupiters are filled in purple, blue and green, respectively. The detection efficiencies in these three regions are close to $100\%$. We also plot the boundaries of the ``desert" identified by \cite{Maz16} with black dashed lines.}
\end{figure*}

\begin{table*}
\centering
\caption{The planetary and stellar parameters for the sample of 295 Kepler planets in 256 systems studied in this work.
The stellar parameters (effective temperature, surface gravity, and metallicity) are derived using the LASP pipeline from LAMOST DR4. The planet radii are revised from Kepler catalogs according to stellar radii calculated based on the LAMOST parameters. }
https://arxiv.org/src/1706.07807v3/anc/hoptune\_tableS1.csv
\end{table*}
\subsubsection*{SI Tables}
\begin{table*}
\centering
\caption{Stellar parameters for the LAMOST-Kepler sample derived using the LASP pipeline from the publicly available LAMOST DR3 ``A, F, G and K type stars catalog'' (dr3.lamost.org/catalogue).
When there are multiple measurements of the same star, the median value is taken and given in this table. Note that the DR3 parameters can have small differences with DR4 parameters used in this work, while they are consistent within uncertainties. The error estimates are based on section 1 of the supporting information of ref. 21.}
https://arxiv.org/src/1706.07807v3/anc/hoptune\_tableS2.csv
\end{table*}

\end{document}